\documentclass[floats,floatfix,aps,prl,showpacs,preprintnumbers,amsmath,amssymb,twocolumn]{revtex4}
\newif\ifpdf\ifx\pdfoutput\undefined\pdffalse\else\pdfoutput=1\pdftrue\fi

\usepackage{graphicx}
\usepackage{graphics}
\usepackage{dcolumn}
\usepackage{bm}

\begin{document}

\title{Pattern formation within \emph{Escherichia coli}: diffusion,
membrane attachment, and self-interaction of MinD molecules}

\author{Rahul V. Kulkarni$^1$}
\email{kulkarni@vt.edu}
\thanks{Present address: Department of Physics, Virginia Polytechnic
and State University, Blacksburg, VA 24061}
\author{Kerwyn Casey Huang$^{1,2}$}
\thanks{Present address: Department of Molecular Biology, Princeton University, Princeton, New Jersey 08540}
\author{Morten Kloster$^1$}
\author{Ned S. Wingreen$^{1,3}$}
\affiliation{$^1$NEC Laboratories America, Inc., Princeton, New Jersey 08540}
\affiliation{$^2$Department of Physics, Massachusetts Institute of Technology, Cambridge, Massachusetts 02139}
\affiliation{$^3$ Department of Molecular Biology, Princeton University,
Princeton, New Jersey 08540}
\date{\today}

\begin{abstract}

In {\it E. coli}, accurate cell division depends upon the oscillation
of Min proteins from pole to pole. We provide a model for the
polar localization of MinD based only on diffusion, a delay for
nucleotide exchange, and different rates of attachment to the bare membrane 
and the occupied membrane. We derive analytically the probability density, 
and correspondingly the length scale, for MinD attachment zones.
Our simple analytical model illustrates the processes giving rise 
to the observed localization of cellular MinD zones.
\end{abstract}

\maketitle

Understanding how proteins are directed to specific locations within
the cell is one of the key challenges of cellular biology. At the
basic level, targeting of proteins to subcellular locations is
governed by {\it physical processes}. A striking example is
the system of Min proteins, which functions as an internal spatial
oscillator 
in {\it E. coli}, and is necessary for accurate cell division
\cite{Yu-99}.
The properties and interactions of the three Min proteins, MinC, D, and
E, have been revealed by recent experiments \cite{deBoer-89,deBoer-92,Bi-93,Bi-91,Huang-96,Hu-02,Raskin-99,Hu-99,Fu-01,Rothfield-03}.
MinD is an ATPase---a protein which binds and hydrolyzes the nucleotide
ATP to ADP. In its ATP-bound form, MinD associates with the inner
membrane 
of the cell \cite{Hu-02,Raskin-99}, where it recruits
cytoplasmic MinC \cite{Huang-96} and MinE \cite{Hu-02} onto the
membrane. Once on the membrane, MinE activates hydrolysis 
of ATP by MinD which results in MinD
dissociating from the membrane. MinE and MinD together produce 
a spatial oscillator with a period of $\sim$40 seconds 
\cite{Hu-02,Raskin-99,Hu-99,Fu-01}. In
each oscillation period, the majority of MinD molecules accumulate at
one end of the cell forming a ``polar zone''. The MinD polar zone then
shrinks toward the end of the cell and a new MinD polar zone forms at
the opposite pole. MinE is observed to form a ring at the medial 
edge of the MinD polar zone \cite{Raskin-97}.
The MinE ring moves along with the polar zone edge as the MinD 
polar zone shrinks toward the end of the cell. 
The dynamics of MinC follows that of MinD \cite{Raskin-99b}.
Complexes of MinC and MinD on the membrane block the formation of
a ring of FtsZ protein \cite{Bi-93}, a necessary first step in determining the 
site of cell division \cite{Bi-91}. The spatial oscillations of MinD and MinC 
from pole to pole thus ensure that an FtsZ ring does not
form near the poles, so that cell division can only occur near midcell. 
Recently, several numerical models have been proposed to explain the 
oscillatory behavior and mechanisms of protein targeting \cite{Howard-01,Meinhardt-01,Kruse-02,Howard-03,Huang-03,Howard-04}.

One of the key emergent properties of the Min oscillations is the
length scale for formation of a new MinD attachment zone. In
filamentous cells ({\it i.e.} cells that grow but do not divide),
several MinD zones are observed in a striped pattern with a
characteristic length separating MinD attachment zones
\cite{Raskin-99,Hale-01}.
In this Letter, we study the processes giving rise to new MinD
attachment zones in the cell. Using a simple model, we demonstrate
analytically that the length separating MinD zones depends on ($i$)
the cytoplasmic diffusion coefficient and the nucleotide-exchange rate
of MinD, and ($ii$) the rate of attachment of ATP-bound MinD to the
membrane.

\begin{figure}
\begin{center}
\resizebox{7.5cm}{!}{\includegraphics{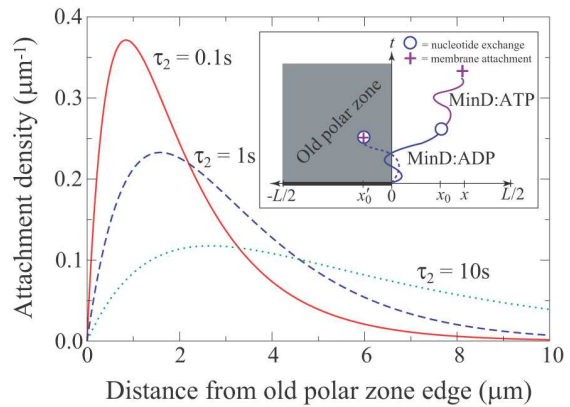}}
\caption{(Color online) Normalized attachment density $\rho(x)/p_{\rm tot}$ (Eq.~\ref{pf-eq}) for
different average attachment waiting times 
$\tau_2$ from 0.1s to 10s, 
with the diffusion coefficient ${\cal D}=2.5\mu$m$^2/$s and
average nucleotide-exchange waiting time
$\tau_1=1$s taken from Ref.~\cite{Huang-03}. 
Inset: cartoon of attachment processes. The solid blue line
shows a MinD:ADP which diffuses to the position $x_0$ before
undergoing nucleotide exchange. The resulting MinD:ATP then continues to
diffuse until it attaches to the bare membrane at $x$, without ever
crossing
the edge of the old polar zone at $x=0$. 
In contrast, the dashed blue line shows a MinD:ADP
which undergoes nucleotide exchange at the position $x_0^\prime$, where it immediately reattaches to the
membrane in the old polar zone. 
\label{times}}
\end{center}
\end{figure}

{\it Model.} Our aim is to illustrate the physical processes of reaction and diffusion that give rise to preferred length scales in the Min-protein system. Our 1D model for these processes 
is abstracted from the 3D numerical model of
Ref.~\cite{Huang-03}, which is based on measured properties of MinD
and MinE and which reproduces the observed pattern in filamentous
cells. Other models have also accounted for the oscillations of the
Min proteins \cite{Meinhardt-01,Howard-01,Kruse-02}, including the
striped pattern seen in filamentous cells. We begin by considering a
fully formed ``old'' polar zone of ATP-bound MinD (MinD:ATP) in one
half of the cell. MinE activates the ATPase activity of MinD, giving
rise to a source of cytoplasmic MinD:ADP. Subsequently, there are
two stages leading up to reattachment of MinD to the membrane: ($i$)
diffusion of each MinD:ADP until nucleotide exchange transforms it
into MinD:ATP, and ($ii$) continued diffusion of the MinD:ATP until it
attaches to the membrane (see inset to Fig.~\ref{times}). The
attachment rate of MinD:ATP increases with the local concentration of
membrane-bound MinD; thus the old polar zone is much stickier than the
bare membrane \cite{Huang-03}. 
Our aim is to derive analytically the density
of attachment $\rho(x)$ of MinD:ATP outside the old polar zone, in
order to obtain the length scale of the new MinD attachment zone.

{\it Filamentous cell, single source.}
We approximate a long, filamentous cell as an infinite 1D line. 
The old polar zone of MinD occupies $x<0$, and bare membrane 
occupies $x>0$ (see inset to Fig.~\ref{times}).
To model the effect of the MinE ring at the edge 
of the MinD polar zone, we first consider that 
MinD:ADP dissociates from the membrane only at $x=0$.
After stage ($i$) -- diffusion of MinD:ADP until nucleotide exchange -- the probability density of each MinD:ATP is 
\begin{equation}
P_1(x) = \int_0^\infty P_{\cal D}(x|t)Q_1(t)dt,\label{p1i-eq}
\end{equation}
where 
$Q_1(t)$ is the waiting-time distribution for single-step 
nucleotide exchange with average waiting time $\tau_1$,
\begin{equation}
Q_1(t) = \frac{1}{\tau_1}e^{-t/\tau_1}, \label{qeq}
\end{equation}
and $P_{\cal D}(x|t)$ is the distribution following diffusion
in 1D for time $t$ with diffusion coefficient ${\cal D}$,
\begin{equation}
P_{\cal D}(x|t) = \frac{1}{\sqrt{4\pi {\cal D}t}}
\exp\left(-\frac{x^2}{4 {\cal D}t}\right).\label{dfeq}
\end{equation}
Substituting Eqs.~\ref{qeq} and \ref{dfeq} into Eq.~\ref{p1i-eq} yields the
initial distribution of MinD:ATP in the cytoplasm
\begin{equation}
P_1(x) = \frac{1}{2\sqrt{{\cal
D}\tau_1}}\exp\left(-\sqrt{\frac{x^{2}}{{\cal D}\tau_1}}\right).\label{p1-eq}
\end{equation}

During stage ($ii$)---diffusion of MinD:ATP until membrane 
reattachment---we initially assume that the region of the 
old polar zone ($x<0$) 
is infinitely sticky for MinD:ATP. Hence, if a MinD:ATP diffuses into
$x<0$, it will immediately reattach and become part 
of the old polar zone. Thus
only those MinD:ATP molecules formed at $x>0$ which {\it
never} cross the origin in the course of cytoplasmic diffusion will
attach to form the new zone. Given an initial position $x_0>0$
of a MinD:ATP, the probability that it diffuses
to some $x>0$ after time $t$ is $P_{\cal D}(x-x_0|t)$. We define the
probability that a MinD:ATP diffuses to some $x>0$ 
after time $t$ {\em without crossing} 
$x=0$ by $P_2(x,x_0|t)$. 
The probability distribution $P_2(x,x_0|t)$ can be calculated using
the continuum version of the Reflection Principle \cite{Feller-68},
\begin{equation}
P_2(x,x_0|t) = P_{\cal D}(x-x_0|t)-P_{\cal D}(x+x_0|t).  
\end{equation}

The attachment density $\rho(x)$ for the bare membrane at $x>0$
can now be obtained from 
\begin{equation}
\rho(x) = \int_0^\infty dx_0\int_0^\infty dt
P_2(x,x_0|t)P_1(x_0)Q_2(t),\label{p-eq}
\end{equation}
where $Q_2(t) = (1/\tau_2) \exp(-t/\tau_2)$ is the distribution
of waiting times for membrane attachment with average waiting time
$\tau_2$. Integrating over $t$ and $x_0$ in Eq.~\ref{p-eq}, we obtain (for $\tau_1\not=\tau_2$)
\begin{equation}
\rho(x) = \frac{p_{\mathrm{tot}}}{\sqrt{{\cal D}\tau_1}- \sqrt{{\cal
D}\tau_2}}\left(e^{-x/\sqrt{{\cal D}\tau_1}}- e^{-x/\sqrt{{\cal
D}\tau_2}}\right),\label{pf-eq}
\end{equation}
where the total probability of attachment to the bare membrane 
at $x > 0$ is
\begin{equation}
p_{\mathrm{tot}} = \frac{\sqrt{\tau_1}}{2 (\sqrt{\tau_1} +
\sqrt{\tau_2})}. \label{rho-eq}
\end{equation}
In the limit $ \tau_{2} \rightarrow 0$, where the
membrane is everywhere perfectly sticky for MinD:ATP, Eq.~\ref{rho-eq}
gives $p_{\mathrm{tot}} = 1/2$, as required by symmetry. In the
opposite limit, $\tau_{2} \gg \tau_{1}$, where sticking of MinD:ATP to
the bare membrane at $x > 0$ is slow, only a small fraction,
$p_{\rm tot} = \sqrt{\tau_1/4\tau_2}$, attach to the bare membrane.

We expect that polymerization of MinD:ATP on the membrane leads to
autocatalytic membrane attachment that enhances the peak in the
attachment density \cite{Huang-03}. Thus, the relevant length scale
to characterize the new MinD:ATP zone is the position of the maximum
in $\rho(x)$, which occurs at 
\begin{equation} x_{\rm{max}} =
\frac{\sqrt{{\cal D}\tau_1\tau_2}}{2
(\sqrt{\tau_2}-\sqrt{\tau_1})}\log\left(\frac{\tau_2}{\tau_1}\right)\label{max1}.
\end{equation}

In Fig.~\ref{times}, we plot the normalized MinD:ATP attachment
density $\rho(x)/p_{\rm tot}$ from Eq.~\ref{pf-eq}.  At large
distances, the profile decays exponentially as expected for a
diffusive process (see Eq.~\ref{p1-eq}).  However, for a simple
diffusive process we expect the profile to have a maximum at the
source position, {\it i.e.}, at $x=0$. In contrast, we find that the
probability of attachment is {\it zero} at the edge of the old polar
zone at $x=0$. This ``zero boundary condition'' follows from the
infinite stickiness of the old polar zone for MinD:ATP. Any MinD:ATP
which forms near the old polar zone has a high probability of crossing
into $x<0$ and immediately reattaching as part of the old polar zone.
Only those MinD:ATP which form sufficiently far from the old polar
zone are likely to reattach as part of the new MinD:ATP attachment
zone.  The two competing effects, the zero boundary condition at $x=0$
and the exponential decay due to diffusion as indicated in
Eq.~\ref{p1-eq}, set the length scale for the formation of the new
MinD:ATP attachment zone. In the model from Ref. \cite{Huang-03}, the
time scale for sticking to a bare membrane is $\tau_2\sim 10$s, giving
$x_{\rm{max}}=2.7\mu$m.  This length scale agrees qualitatively with
half the center-to-center distance ($\sim 3\mu$m) between neighboring
MinD attachment zones as observed in Ref. \cite{Raskin-99}.

{\it Filamentous cell, distributed source.}  In the preceding
analysis, the polar-zone edge at $x=0$ was taken to be the only source
of MinD:ADP.  However, in experiments some MinE is observed throughout
the MinD polar zone {\cite{Raskin-97,Rothfield-03}, suggesting that
cytoplasmic MinD:ADP is released throughout the old polar zone as
well.  What effect, if any, does this have on the length scale for
formation of the new MinD attachment zone? Instead of assuming a
single source for MinD:ADP at $x=0$, we now consider a source with
distribution $w(x_s)$, $x_s \le 0$. This distributed source of
MinD:ADP modifies Eq.~\ref{p1i-eq}:
\begin{equation}
P_1(x) \rightarrow \int_0^\infty dt \int_{-\infty}^0dx_s w(x_{s})
P_{\cal D}\left(x-x_s|t\right)Q_1(t),\label{p1mod-eq}
\end{equation}
where $x_s$ describes the position of the MinD:ADP when it leaves the 
membrane. Integrating over $t$, we find for $x > 0$,
\begin{equation}
P_1(x)=\frac{1}{2\sqrt{{\cal D}\tau_1}}e^{-x/\sqrt{{\cal D}\tau_1}}
\left[\int_{-\infty}^0dx_s w(x_{s}) e^{x_s/\sqrt{{\cal D}\tau_1}}\right].
\end{equation}
Comparing to Eq.~\ref{p1-eq}, we see that a distributed source of MinD:ATP 
simply reduces $P_{1}(x)$ by the constant factor in square brackets. 
Therefore, the attachment 
density is the same as in Eq.~\ref{pf-eq}, with $p_{\rm tot}$ reduced by the
constant factor in square brackets. This leaves $x_{\rm{max}}$
unchanged,
so the length scale for formation of the new MinD attachment zone is
identical.

{\it Finite polar-zone attachment probability.} How is the
reattachment profile altered by relaxing the approximation that the
old polar zone is infinitely sticky for MinD:ATP?  To treat the case
where the mean attachment time to the old polar zone is $\tau_3>0$, we
introduce the quantities $\rho_{c}(x,t)$ and $\rho_{m}(x,t)$,
respectively denoting the densities of cytoplasmic- and membrane-bound
MinD:ATP as functions of time. Our previous expression for the
membrane attachment density $\rho(x)$ (Eq.~\ref{pf-eq}) is then
equivalent to $ \rho_{m} (x,t\rightarrow\infty)$ for the case
$\tau_3 = 0$ and $x >0$.

\begin{figure}
\begin{center}
\resizebox{7.5cm}{!}{\includegraphics{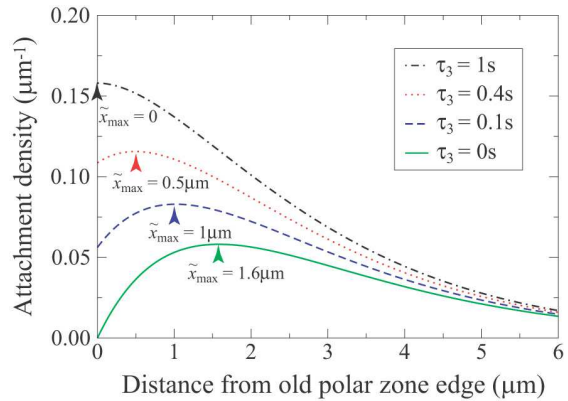}}
\caption{(Color online) Attachment density $\tilde{\rho}(x)$ (Eq.~\ref
{Fneweq}) for different values of $\tau_3$, the 
average attachment time of MinD:ATP 
in the old polar zone, with ${\cal D}=2.5\mu$m$^2/$s, 
average nucleotide exchange waiting time  
$\tau_1=1s$, and 
average bare-membrane attachment time
$\tau_2=1s$. Note the shift of the maximum toward $x=0$ as $\tau_3\rightarrow \tau_2$. 
\label{fig2}}
\end{center}
\end{figure}

\begin{figure*}
\begin{center}
\resizebox{14cm}{!}{\includegraphics{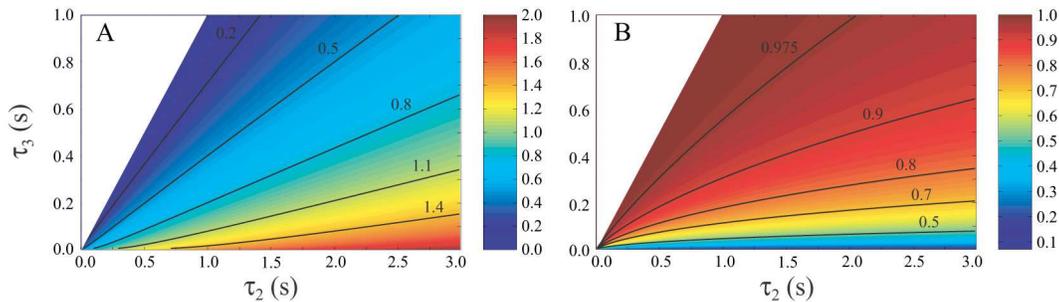}}
\caption{(Color) (a) The position of the 
maximum in the MinD-attachment profile, $\tilde{x}_{\rm max}$ (Eq. \ref{xmaxneweq}), as a function of $\tau_2\in [0\mbox{s},3\mbox{s}]$ and $\tau_3\in [0\mbox{s},1\mbox{s}]$.  
(b) The ratio of densities $f(0)/f(\tilde{x}_{\rm max})$.
\label{fig3}}
\end{center}
\end{figure*}

The time-dependent densities of MinD in the cytoplasm and on the membrane
are described by the following equations
\begin{eqnarray}
\frac{\partial \rho_{c}(x,t)}{\partial t} & = &{\cal D} \frac{\partial^{2} \rho_{c} (x,t) }{\partial x^2} - \frac{1}{\tau(x)} \rho_{c}(x) \\
\frac{\partial \rho_{m}(x,t)}{\partial t}& = & \frac{1}{\tau(x)} \rho_{c}(x,t) \label{integrateeq}
\end{eqnarray}
where
\begin{equation}
\tau(x) = \left\{\begin{array}{cc}
\tau_2 \;\;\;\; & x >0  \\
\tau_3 \;\;\;\; &   x \leq 0
\end{array}\right.
\end{equation}
We are interested in 
the final density of MinD on the membrane
$\rho_{m}(x,\;t\!\rightarrow\!\infty)$  for $x > 0$,
which, from integrating Eq.~\ref{integrateeq}, is given by $\frac{1}{\tau_2} 
\int_{0}^{\infty} \rho_{c}(x,t) dt \equiv f(x)/{\tau_2}$. 
Similarly, we define $\rho_{m}(x,\;t\!\rightarrow\!\infty) \equiv f(x)/{\tau_3}$ for $x \leq 0$. Then 
the unknown function  
$f(x)$ satisfies 
\begin{equation}
\rho_{c} (x,0) = - {\cal D} \frac{d^{2} f}{d x^{2}} + \frac{1}{\tau(x)} f(x).\label{f-eq}
\end{equation}
Note that $\rho_{c}(x,0)$ is the initial distribution of MinD:ATP 
in the cytoplasm,
which is given by Eq.~\ref{p1-eq}. We can now readily solve
Eq.~\ref{f-eq} for $f(x)$ in the regions $x \leq 0$ and $x > 0$,
subject to the boundary conditions that $f(x)$ vanishes at $\pm
\infty$ and is continuous and differentiable at $x=0$.

The new MinD:ATP
attachment profile for $x>0$ with a finite binding rate $1/\tau_3$  in the 
old polar zone is
\begin{eqnarray}
\lefteqn{\tilde{\rho}(x) = \frac{f(x)}{\tau_2} = 
\frac{1}{2(\tau_1-\tau_2)}\sqrt{\frac{1}{{\cal D}\tau_2}} \left( \sqrt{\tau_1\tau_2}e^{-x/\sqrt{{\cal D}\tau_1}}\right. }\nonumber\\
&&\left.- \frac{2\tau_2\sqrt{\tau_1\tau_3}+2\tau_2\tau_3+\tau_1\tau_2-\tau_1\tau_3}{(\sqrt{\tau_1}+\sqrt{\tau_3})(\sqrt{\tau_2}+\sqrt{\tau_3})}e^{-x/\sqrt{{\cal D}\tau_2}}\right). \label{Fneweq}
\end{eqnarray}
In Fig.~\ref{fig2}, we plot $\tilde{\rho}(x)$ for $\tau_2=1$s and four
values of $\tau_3$: 0s, 0.1s, 0.4s, and 1s. For $\tau_3=0$s, the density
profile in Eq.~\ref{pf-eq} is reproduced. For $\tau_3=\tau_2=1s$, the
profile is symmetric about $x=0$ with a maximum at $x=0$. The intermediate
values show a shift in the maximum toward $x=0$ as $\tau_3$ approaches $\tau_2$, and a finite density at $x=0$. 

The maximum occurs at
\begin{equation}
\tilde{x}_{\rm max}=\frac{\sqrt{{\cal D}\tau_1\tau_2}}{\sqrt{\tau_2}-
\sqrt{\tau_1}}\log\left(\frac{\tau_2(\sqrt{\tau_1}+
\sqrt{\tau_3})(\sqrt{\tau_2}+\sqrt{\tau_3})}{2\tau_2\sqrt{\tau_1\tau_3}+
2\tau_2\tau_3-\tau_1\tau_3+\tau_1\tau_2}\right), \label{xmaxneweq}
\end{equation}
which is less than $x_{\rm max}$ (see Eq.~\ref{max1}) for all nonzero $\tau_3$.
Another quantity of interest is the ratio of 
the MinD:ATP density at the polar-zone edge to the peak density,
$\eta=f(0)/f(x_{\rm max})$. This is a measure of
the spatial definition of the new polar zone.
In Fig.~\ref{fig3}, we plot (a) $\tilde{x}_{\rm max}$ and (b)
$f(0)/f(\tilde{x}_{\rm max})$ for $\tau_2\in [0\mbox{s},3\mbox{s}]$ and $\tau_3\in [0\mbox{s},1\mbox{s}]$,
$\tau_3<\tau_2$. Interestingly, except for very small $\tau_3$,
$\tilde{x}_{\rm max}$ is a function only of the ratio $\tau_2/\tau_3$.
Note also that the ratio $\eta$ is 
small only for very low values of $\tau_3$.  
These results demonstrate that the asymmetry of
binding probabilities
to the old polar zone and to the bare membrane is crucial 
to establish a new attachment zone well separated from the old one.

In summary, we have studied the processes that give rise to the length
scale observed in cellular MinD attachment zones. Our analysis
indicates that this length scale is an emergent property of the system
which depends primarily on ($i$) the rate of nucleotide exchange, ($ii$)
the rates of membrane attachment, and ($iii$) the diffusion constant for
MinD. It is important to note that these are biochemical properties of
MinD which can be fine tuned by evolution. Our analysis suggests that
these parameters are tuned to block minicelling at the cell poles and
at the same time to permit cell division at the 
cell
center. Finally, we
note that the Min system serves
as an example of how physical processes can produce 
subcellular protein localization. While MinD is
``targeted'' to the cell poles, the mechanism is not based on
recruitment by a biological target, but rather on a dynamical
physical
instability in the system.

We thank Kevin Beach for valuable suggestions. Partial funding for
this research was provided by the MRSEC program of the NSF under grant
number DMR-0213282.

\end{document}